\documentclass[showpacs,floatfix,twocolumn,byrevtex,superscriptaddress]{revtex4-1}
\usepackage[english]{babel}
\usepackage{amsmath}
\usepackage{graphicx}
\usepackage{float}
\usepackage{bm}

\usepackage{ulem}
\usepackage[dvipsnames,usenames]{color}

\newcommand{\icm}{\ensuremath{~\textrm{cm}^{-1}}}% % cm-1

\begin{document}

\bibliographystyle{apsrev}

\title{Optical Conductivity Evidence of Clean-Limit Superconductivity in LiFeAs
%Optical Conductivity of Superconducting LiFeAs \\
%Clean-limit Superconductivity in LiFeAs \\
%Optical conductivity of a clean-limit superconductor LiFeAs \\
%Optical evidence of strongly correlated superconductivity in LiFeAs
}

\author{R. P. S. M. Lobo}
\email[]{lobo@espci.fr}
\affiliation{ESPCI-ParisTech, PSL Research University; CNRS, UMR 8213; Sorbonne Universit\'es, UPMC Univ. Paris 6; LPEM, 10 rue Vauquelin, F-75231 Paris Cedex 5, France}
%\affiliation{LPEM, ESPCI-ParisTech, PSL Research University, 10 rue Vauquelin, F-75231 Paris Cedex 5, France}
%\affiliation{, F-75005 Paris, France}
%\affiliation{, F-75005 Paris, France}

\author{G. Chanda}
\affiliation{Dresden High Magnetic Field Laboratory (HLD-EMFL),
Helmholtz-Zentrum Dresden-Rossendorf, D-01314 Dresden, Germany}
\affiliation{Institut f\"{u}r Festk\"{o}rperphysik, Technische
Universit\"{a}t Dresden, D-01062 Dresden, Germany}

\author{A. V. Pronin}
\affiliation{Dresden High Magnetic Field Laboratory (HLD-EMFL),
Helmholtz-Zentrum Dresden-Rossendorf, D-01314 Dresden, Germany}
\affiliation{A. M. Prokhorov Institute of General Physics,
Russia Academy of Sciences, 119991 Moscow, Russia}
\affiliation{1. Physikalisches Institut, Universit\"{a}t Stuttgart,
Pfaffenwaldring 57, D-70550 Stuttgart, Germany }

\author{J. Wosnitza}
\affiliation{Dresden High Magnetic Field Laboratory (HLD-EMFL),
Helmholtz-Zentrum Dresden-Rossendorf, D-01314 Dresden, Germany}
\affiliation{Institut f\"{u}r Festk\"{o}rperphysik, Technische
Universit\"{a}t Dresden, D-01062 Dresden, Germany}

\author{S. Kasahara}
\affiliation{Department of Physics, Kyoto University, Kyoto 606-8502, Japan}
%\affiliation{Research Center for Low Temperature and Materials Sciences, Kyoto University, Kyoto 606-8501, Japan}

\author{T. Shibauchi}
\affiliation{Department of Physics, Kyoto University, Kyoto 606-8502, Japan}
\affiliation{Department of Advanced Materials Science, University of Tokyo, Chiba 277-8561, Japan}

\author{Y. Matsuda}
\affiliation{Department of Physics, Kyoto University, Kyoto
606-8502, Japan}

\date{\today}
\begin{abstract}
We measured the optical conductivity of superconducting LiFeAs. In
the superconducting state, the formation of the condensate leads to
a spectral-weight loss and yields a penetration depth of 225~nm. No
sharp signature of the superconducting gap is observed. This
suggests that the system is likely in the clean limit. 
A Drude-Lorentz parametrization of the data in the normal state reveals 
a quasiparticle scattering rate supportive of spin fluctuations and proximity to
a quantum critical point.
\end{abstract}
\maketitle

%%%%%%%%%%%%%%%%%%%%%%%%%%%%%%%%%%%%%%%%%%%%%%%%%%%%%%%%%%%%%%%%%%%%%%%%%%%%%%%
%
% Introduction
%
\section{Introduction}

Unlike iron-based superconductors of other families,
superconductivity, rather than magnetic order, emerges in LiFeAs
at zero chemical doping \cite{WangLiFeAs, TappLiFeAs}. The
superconducting transition temperature is rather high ($T_{c} = 18$
K at ambient pressure) and the electronic structure of LiFeAs
shows no signatures of good nesting \cite{BorisenkoLiFeAs}.
Angle-resolved photoemission spectroscopy (ARPES) \cite{BorisenkoLiFeAs, UmezawaLiFeAs},
scanning tunneling spectroscopy (STM) \cite{Chi2012},
neutron-scattering \cite{InosovLiFeAs},
penetration-depth \cite{KimLiFeAs, HashimotoLiFeAs}, and
thermal-conductivity \cite{TanatarLiFeAs} measurements suggest
multiband superconductivity and full in-plane gaps with no
indications of nodes and with either absent \cite{InosovLiFeAs,
TanatarLiFeAs, Min2013} or modest \cite{BorisenkoLiFeAs,
UmezawaLiFeAs, AllanLiFeAs} gap anisotropy.

Usually, LiFeAs samples have a very large residual resistivity
ratio. This is a general indication of high sample quality
\cite{KasaharaLiFeAs, HeyerLiFeAs, Rullier2012}. As superconductivity appears in
the stoichiometric composition, the properties of LiFeAs are not influenced
by doping-induced defects and impurities. The availability of LiFeAs in
form of large high-quality single crystals, makes it a prime material
for optical investigations.

To date, only \citeauthor{Min2013} \cite{Min2013} reported on the
far-infrared conductivity of LiFeAs. They described their data in
the framework of multiple superconducting gaps, and found two
fully open, gaps at $2\Delta_0 = 3.2$ meV and $2 \Delta_0 = 6.3$ meV. These data
were further analyzed in terms of Eliashberg theory \cite{Hwang2015}.
STM also finds two homogeneous nodeless gaps, but with values twice as
large \cite{Chi2012}.

In this paper, we study the optical properties of superconducting
LiFeAs. The overall response of this material is similar to other
FeAs-based superconductors. Upon entering the superconducting state,
LiFeAs shows a loss of spectral weight related to the formation of
the superconducting condensate with a penetration depth of 225~nm.
However, contrary to \citeauthor{Min2013} findings \cite{Min2013}, we do not
observe a clear gap signature, suggesting that the system is likely in
the clean limit. The high residual-resistance ratio of our sample and
the presence of quantum oscillations in samples from the same 
batch \cite{Putzke2012} further support this clean limit picture.
In the normal state, a Drude-Lorentz decomposition of the optical
conductivity leads to a scattering rate that evolves linearly with
temperature, a property observed in other optimally doped pnictide
superconductors and suggestive of the proximity to a quantum critical
point (QCP).

%%%%%%%%%%%%%%%%%%%%%%%%%%%%%%%%%%%%%%%%%%%%%%%%%%%%%%%%%%%%%%%%%%%%%%%%%%%%%%%
%
% Methods
%
\section{Methods}

High-quality single crystals of LiFeAs were grown by a self-flux
method using Li ingots and FeAs powder. The starting materials were
placed in a BN crucible, and sealed in a quartz tube. The tube was
heated to 1100$^{\circ}$C, then slowly cooled down to 600$^{\circ}$C. The typical linear size of the single crystals obtained
was 3 to 5 mm in each direction. DC resistivity measurements, showing 
$T_c = 18$~K, have been reported earlier \cite{KasaharaLiFeAs}.

Near normal incidence reflectivity from 20 to $6\,000\icm$ was
measured on Bruker IFS113 and IFS66v spectrometers at 5, 10, 15, 20,
25, 50, 100, 200, and 300 K. The absolute reflectivity of the sample
was obtained with an \textit{in situ} gold overfilling technique \cite{Appl.Opt.32.2976}.
The reflectivity has an absolute accuracy
better than 0.5\% and a relative accuracy better than 0.1\%. The
data were extended to the visible and UV ($5\,000$ to $40\,000\icm$)
at room temperature with an AvaSpec-$2048 \times 14$ spectrometer.

LiFeAs is highly air sensitive. To preserve the sample, it was
kept in a sealed vial in Ar atmosphere and mounted in
the cryostat cold finger inside a glove box, also in Ar atmosphere.
The sample was cleaved prior to each temperature run.

%%%%%%%%%%%%%%%%%%%%%%%%%%%%%%%%%%%%%%%%%%%%%%%%%%%%%%%%%%%%%%%%%%%%%%%%%%%%%%%
%
% Results
%
%
\section{Results}

\begin{figure}[htb]
  \includegraphics[width=0.9\columnwidth]{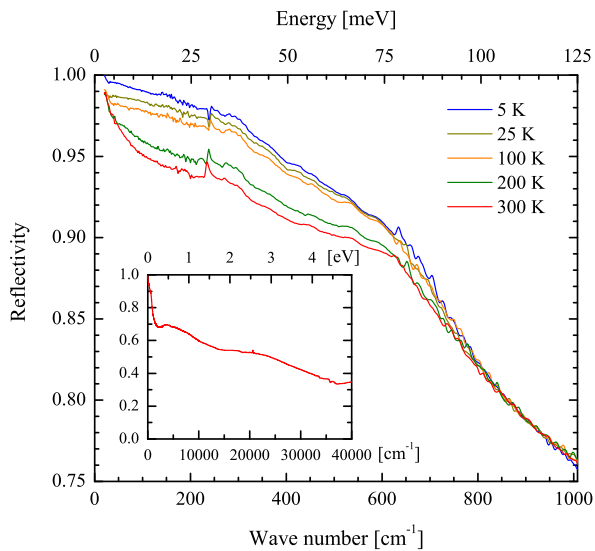}
\caption{(color online) In-plane infrared reflectivity of LiFeAs.
 The inset shows the reflectivity at 300 K up to 5 eV.}
  \label{Fig1}
\end{figure}

Figure \ref{Fig1} displays the \textit{ab}-plane reflectivity of
LiFeAs above and below $T_c$. The low-frequency reflectivity
increases steadily with decreasing temperature. Contrary to an
s-wave BCS superconductor, there is no sharp rise of the
reflectivity upon entering the superconducting state. There is also
no sign of a flat 100\% reflectivity at low frequencies. The sharp
peak around 240\icm (30 meV) is a polar phonon of LiFeAs. 

The real part, $\sigma_{1}(\omega)$, of the optical conductivity was
derived from the reflectivity through Kramers-Kronig analysis. At
low frequencies we utilized either a Hagen-Rubens ($1 - A
\sqrt{\omega}$) or a superconducting ($1 - A \omega^4$)
extrapolation. At high frequencies we applied a constant
reflectivity to 12.5 eV followed by a $\omega^{-4}$ free-electron
termination.

\begin{figure}[htb]
  \includegraphics[width=0.9\columnwidth]{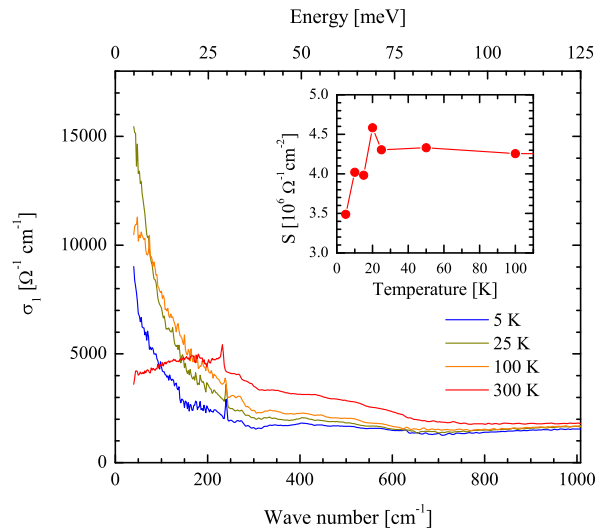}
\caption{(color online) Optical conductivity of LiFeAs above and
below $T_c$. The inset shows the spectral weight below 100 K calculated by 
integrating $\sigma_1$ up to 0.5 eV. Below $T_c$, it 
shows a drop related to the formation of the superconducting condensate.}
  \label{Fig2}
\end{figure}

Figure \ref{Fig2} shows $\sigma_1(\omega)$ at various temperatures 
for wave numbers above 40\icm, our limit of confidence in Kramers-Kronig obtained data.
At 300 K, the optical conductivity signals an almost incoherent transport:
it depicts a very broad Drude-like peak as well as a bump around 200\icm (25
meV), which could be related to low-energy interband transitions, in
particular in view of the presence of shallow bands in
LiFeAs \cite{BorisenkoLiFeAs}. However, we cannot exclude a small
surface contamination due to the fragile chemical stability of
LiFeAs. We will not discuss this feature any further. When cooling
down the sample, in the normal state, the Drude-like term increases
and fully develops into a coherent peak. Upon crossing into the
superconducting state, the low frequency optical conductivity decreases
for energies below 600\icm (75 meV). 
The inset of Fig.~\ref{Fig2} shows a sum-rule analysis that is discussed
in Sec.~\ref{PenDepth}.
%

%%%%%%%%%%%%%%%%%%%%%%%%%%%%%%%%%%%%%%%%%%%%%%%%%%%%%%%%%%%%%%%%%%%%%%%%%%%%%%%
%
% Clean Limit Superconductivity
%
\section{Clean Limit Superconductivity}

Looking at Fig.~\ref{Fig2}, one cannot pinpoint a clear-cut sharp energy edge
characteristic of an s-wave, BCS superconducting gap. There are two
main possibilities for the absence of a superconducting gap edge.
The first option is a strong anisotropy in the gap, preferably with
nodes. Although some anisotropy has been observed by
ARPES \cite{BorisenkoLiFeAs, UmezawaLiFeAs} and momentum-resolved
tunneling spectroscopy \cite{AllanLiFeAs}, it is too small to account
for a non-vanishing low-energy optical conductivity originating from unpaired
quasiparticles. Indeed, virtually every measurement of the gap in
LiFeAs indicates a nodeless state \cite{BorisenkoLiFeAs, Chi2012,
UmezawaLiFeAs, InosovLiFeAs, KimLiFeAs, HashimotoLiFeAs,
TanatarLiFeAs, Min2013, AllanLiFeAs}.

The second possibility is a superconductor in a clean limit, which
seems to be the case for LiFeAs \cite{TanatarLiFeAs, KasaharaLiFeAs,
InosovLiFeAs, KimLiFeAs}. As pointed out by
\citeauthor{Kamaras1990} \cite{Kamaras1990}, one does not see the gap
in the optical conductivity of a clean superconductor. To understand
this statement, let us remark that, optically speaking, the clean
limit corresponds to the case where the quasiparticle scattering
rate ($\tau^{-1}$) is small compared to the superconducting gap,
\textit{i.e.}, $\tau^{-1} < 2 \Delta$. In the Drude model, most of the spectral
weight is comprised below $\tau^{-1}$. At the superconducting transition, spectral
weight below $2 \Delta$ is transferred to a $\delta(\omega)$ function
representing the condensate. If $2
\Delta$ is larger than $\tau^{-1}$ the spectral weight around $\omega = 2
\Delta$ is vanishingly small --- both in the normal and superconducting states ---
and thus no clear signature of the gap in
the optical conductivity exists.
Utilizing the Drude-Lorentz model
discussed further in this paper 
and constraining our fits to respect the temperature 
dependence of the resistivity, 
we estimate the low-temperature
scattering rate to be close to 3~meV (25\icm). Several estimates for
the gap value in LiFeAs are available. 
The smallest value reported
is 3~meV ($2 \Delta = 25\icm$) \cite{Min2013}. Every other estimate
and measurement fall into the range between 4~meV ($32\icm$) and
10~meV ($80\icm$) \cite{InosovLiFeAs, BorisenkoLiFeAs, Chi2012,
UmezawaLiFeAs}. This strongly supports the picture where the absence
of a gap signature in the optical conductivity of LiFeAs is a clean-limit effect.

Another confirmation of the clean-limit superconductivity in
LiFeAs comes from the analysis of the frequency-dependent
penetration depth,
$\lambda(\omega)=[\mu_0 \, \omega \, \sigma_{2}(\omega)]^{-1/2}$,
where $\mu_0$ is the vacuum permeability. For a BCS superconductor of
arbitrary scattering, the measured penetration depth (as defined
above) is related to the London penetration depth
($\lambda_{L}=c/\omega_{p}$, where $\omega_{p}$ is
the free-carrier plasma frequency) in a complex way, with the
mean free path being involved in the relation \cite{Miller1959}. In
terms of optical response, one can show that at frequencies above
the scattering rate (but still well below $\omega_{p}$),
$\lambda(\omega)$ coincides with
$\lambda_{L}$. This is because the penetration depth at a fixed
frequency is defined by the integral response of the carriers at
all frequencies below this one. At high enough frequencies
(above the scattering rate) the unpaired electrons screen the
external field as efficiently as the superconducting currents. Note that at
frequencies above the scattering rate, the unpaired electrons
do not scatter. In the clean
limit at $T=0$, there are no unpaired electrons at all, thus the
high-frequency value of the penetration depth (the London
penetration depth) will span down to zero frequency. Conversely,
in the dirty limit, there are some unpaired electrons even at $T=0$.
The response of these electrons appear in $\sigma_{1}(\omega)$ at
frequencies between the gap value and the scattering rate. In
this frequency range, the measured penetration depth will deviate
from $\lambda_{L}$. The $\lambda_{L}$ value will not be recovered
at $\omega=0$ because a part of electrons remains unpaired and do
not participate in the field screening [but instead contribute to
$\sigma_{1}(\omega)$].

\begin{figure}[htb]
  \includegraphics[width=0.9\columnwidth]{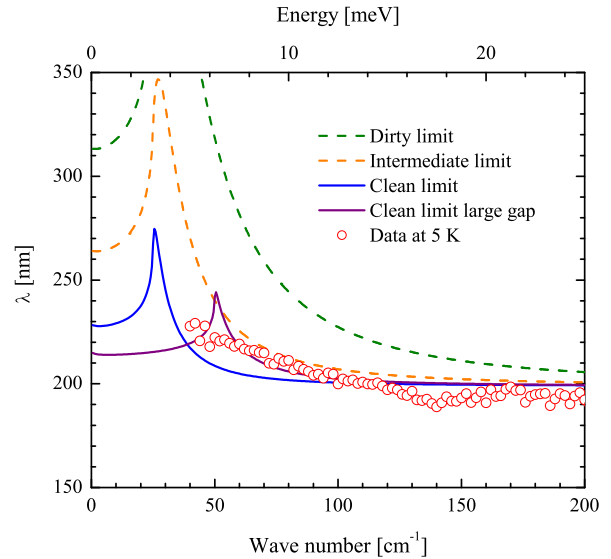}
\caption{(color online) Penetration depth of LiFeAs at 5 K
(symbols). The circles are the data and the lines show BCS calculations 
for the penetration depth. The ``Dirty limit'' 
simulation has  $2 \Delta = 25\icm$ and $\tau^{-1} = 50\icm$. ``Intermediate''
and ``Clean limit'' calculations utilize the same value for the gap and $\tau^{-1}$ 
of 25\icm and 10\icm, respectively. An optical gap of 50\icm\ together with a 10\icm\
scattering rate produce the ``Clean limit large gap'' simulation.}
  \label{Fig3_lambda}
\end{figure}

Figure \ref{Fig3_lambda} shows that, at our lowest temperature,
the spectrum of $\lambda(\omega)$ is basically flat below $600\icm$. 
According to the above, this
indicates that our sample is in the clean limit. 
For comparison,
we utilized the parametrization of
\citeauthor{Zimmermann} \cite{Zimmermann} to calculate the 
frequency dependence for a BCS superconductor
with different combinations of optical gap and scattering rate values. 
Regardless of the gap value, our data correspond to clean limit calculations.

%%%%%%%%%%%%%%%%%%%%%%%%%%%%%%%%%%%%%%%%%%%%%%%%%%%%%%%%%%%%%%%%%%%%%%%%%%%%%%%
%
% Spectral Weight Transfer and Penetration Depth
%
\section{Sum-Rule Penetration Depth}
\label{PenDepth}

Rather than the superconducting gap,
the important optical signature of the superconducting
transition is the loss of spectral weight below $T_c$, related to
the formation of the superfluid condensate. The inset of
Fig.~\ref{Fig2} shows the spectral weight ($S$) corresponding to an
integration of the optical conductivity up to $4\,000\icm$ (0.5 eV), below 100 K.
In the normal state, this value is constant up to 300 K, within 4\%.
Below $T_c$, the occurrence of the superfluid condensate implies a
transfer of spectral weight from finite frequencies to a
$\delta(\omega)$ function representative of the infinite DC
conductivity. As the measured real part of the optical conductivity
has no access to zero frequency, the value of its integral drops
when the superfluid forms. 
The difference between the spectral weights in the normal
and superconducting states leads to the penetration depth through 
$\lambda^2 = \pi \varepsilon_0 c^2 / [2 (S_n - S_{sc})]$, where $\varepsilon_0$ is the vacuum permittivity and the subscripts in $S$ refer to
the normal and superconducting states. For LiFeAs, we find $\lambda
= 225$~nm at 5 K. This is in very close agreement to values obtained
from neutron-scattering (210 nm) \cite{InosovLiFeAs}, infrared (240
nm) \cite{Min2013}, transport (210 nm) \cite{KasaharaLiFeAs}, and muon
spin rotation (195 nm) \cite{PrattLiFeAs} data, as well as with our
calculations from $\sigma_2(\omega)$, discussed above and shown in
Fig.~\ref{Fig3_lambda}.

%%%%%%%%%%%%%%%%%%%%%%%%%%%%%%%%%%%%%%%%%%%%%%%%%%%%%%%%%%%%%%%%%%%%%%%%%%%%%%%
%
% Normal State Scattering
%
\section{Normal State Scattering}

\begin{figure}[tb]
  \includegraphics[width=0.9\columnwidth]{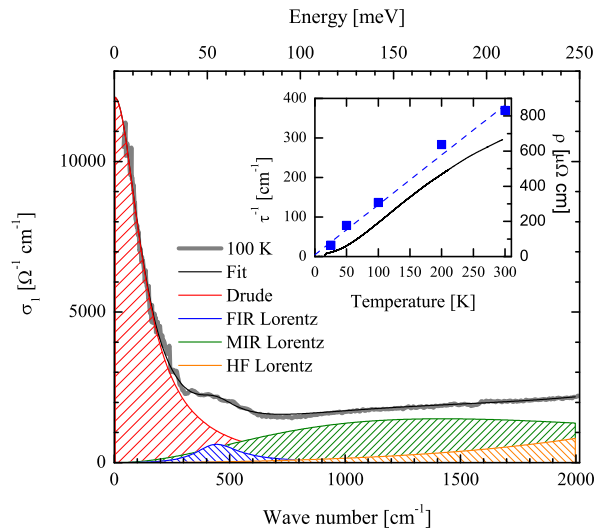}
\caption{(color online) Drude-Lorentz modeling of the optical
conductivity of LiFeAs at 100 K. The thick gray line is the data and
the thin black line is a fit with a single Drude term 
($\Omega_p = 9980\icm$ and $\tau^{-1} = 137\icm$)
and 4 Lorentz terms [far-infrared (FIR), mid-infrared (MIR) and 2 for
higher frequencies (HF)]. Individual contributions are shown as
hatched areas. The inset shows the temperature dependence of the DC 
resistivity (solid line); the Drude
term scattering rate (squares) and a linear fit to the
$\tau^{-1}(T)$ data (dashed line).}
  \label{Fig4}
\end{figure}

Above $T_{c}$, $\sigma_{1}(\omega)$ shows a metallic response which
can be modelled by a Drude-Lorentz optical conductivity:
\begin{equation}
\sigma_{1}(\omega)=\varepsilon_0\left[\frac{\Omega^{2}_{p}}
{\tau(\omega^{2}+\tau^{-2})}+\sum_k\frac{\gamma_k\omega^{2}S_k^{2}}
{(\Omega_k^{2}-\omega^{2})^{2}+\gamma_k^{2}\omega^{2}}\right].
\label{eq.1}
\end{equation}
The first term in Eq.~\ref{eq.1} corresponds to a free-carrier Drude
response, characterized by a plasma frequency ($\Omega_{p}$) and a
scattering rate ($\tau^{-1}$). The second term is a sum of Lorentz
oscillators characterized by a resonance frequency ($\Omega_k$), a
line width ($\gamma_k$), and a plasma frequency ($S_k$). Figure
\ref{Fig4} shows the result of a fit with Eq.~\ref{eq.1} to our data at 100 K. We
utilized a Drude peak (hatched red area) to describe the free
carriers and 4 Lorentz terms to account for transitions in the 
infrared. These 5 contributions
account for the optical conductivity at all temperatures above $T_c$
up to $6\,000\icm$ (0.75 eV). The scattering rate of the Drude
term is the only parameter with a significant temperature dependence. Our Drude fitting parameters were constrained in order to have a temperature dependence of $\sigma_1(0)$ following the inverse DC resistivity.

Although band-structure calculations predict multiple bands at the
Fermi level \cite{NekrasovLiFeAs, SinghLiFeAs}, one has to consider
that the optical conductivity is a reciprocal-space averaged
quantity. Therefore, all bands with similar carrier lifetimes will
contribute to the same Drude term in $\sigma_1$. In this
perspective, \citeauthor{PhysRevB.81.100512}
\cite{PhysRevB.81.100512} showed that two Drude terms (a narrow one
with a small scattering rate and a broad one with a large scattering
rate) are sufficient to describe the optical conductivity of most
iron-arsenide superconductors. There are two important points to
remark in this double Drude fitting: (i) most, if not all, of the
temperature dependence of the spectra is related to the narrow Drude
peak, in particular to its scattering rate; and (ii),
the large scattering rate Drude term systematically leads to a mean
free path comparable to the lattice parameter. Therefore, the broad
Drude term is representative of an incoherent conductivity, probably
with bound carriers. In this case, it can be substituted by a
Lorentz oscillator with the proper spectral weight. We chose a
low-energy Lorentz approach for our fit. This choice is
substantiated by the fact that, had we utilized a broad Drude term
in our fits, its scattering rate would be around $2\,500\icm$ (0.3
eV). Taking a Fermi velocity of $\sim 0.4$~eV\AA\ \cite{Ding2011}, one
would find a mean free path in the range of 1~\AA, which is
obviously smaller than the unit cell size. Looking at mobility
values reported for LiFeAs \cite{KasaharaLiFeAs}, one can safely
assume that the narrow, coherent, Drude peak is representative of
the electron bands.

A multiband analysis of transport data by \citeauthor{Rullier2012} \cite{Rullier2012}, 
proposes a quadratic temperature dependence for the electron and hole scattering rates 
in LiFeAs. Their $T^2$ coefficients are very large, suggestive of strong spin fluctuations.
The inset of Fig.~\ref{Fig4} shows the temperature dependence of the
Drude scattering rate, obtained by fitting the data in the normal
state with Eq.~\ref{eq.1}. We do not have enough temperatures to make a 
strong claim, but our data seem to be better described by a linear
temperature dependence (dashed line), a trend that is also compatible with 
spin fluctuations. Interestingly, the actual resistivity of LiFeAs (solid line) shows 
neither a linear, nor a quadratic temperature evolution of $\tau^{-1}$. 
This fact is related to the multiband 
character of pnictides. Electron-like and hole-like transport properties
vary differently with temperature. At low temperatures electron carriers have a dominating role
whereas a crossover regime appears at higher temperatures where electrons and holes
have similar mobilities. This seems to be a common trend in optimally doped FeAs-based
superconductors \cite{Dai2013}. 

The multiband character of the transport properties, both DC and optical,
suggests a material with strong spin fluctuations associated with a 
competition between magnetism and superconductivity
and a possible existence of a quantum critical point. The ground
state of (nominally) undoped LiFeAs is a superconductor with no
long-range magnetic order. This suggests that spin fluctuations are
not important in this material. Indeed,
\citeauthor{BorisenkoLiFeAs} \cite{BorisenkoLiFeAs} interpreted their
ARPES data in the framework of no static or fluctuating magnetic
order. This observation would be at odds with the spin-fluctuations
driven superconductivity leading to an $s_{\pm}$ order parameter
proposed for iron pnictides in
general \cite{Mazin2008,Cvetkovic2009}, and for LiFeAs in
particular \cite{UmezawaLiFeAs}. However, a different picture emerges
from NMR experiments. \citeauthor{Ma2010} \cite{Ma2010} showed that
small deviations from stoichiometry can tune LiFeAs from a material
with spin fluctuations to a system with a spin-density-wave QCP.
They assign these deviations to the easiness of reversibly
intercalating lithium in interstitial sites. This QCP gets further
support from the linear temperature dependence of the scattering
rate shown in Fig.~\ref{Fig4}.

%%%%%%%%%%%%%%%%%%%%%%%%%%%%%%%%%%%%%%%%%%%%%%%%%%%%%%%%%%%%%%%%%%%%%%%%%%%%%%%
%
% Conclusions
%
\section{Summary}

In summary, 
our optical results show a consistent picture of clean-limit superconductivity in LiFeAs.
In the normal state
LiFeAs shows an optical conductivity dominated by a
narrow Drude-like peak with a strong temperature dependence. When
crossing into the superconducting state, this Drude-like peak shows
a gradual decrease in its spectral weight, characteristic of a superfluid
condensate. From the lost spectral weight, we calculate a
penetration depth of 225~nm at 5 K. We did not observe any sharp
edge in the spectra, indicating that no superconducting gap
signature is observed in the infrared. We conclude that this is a
consequence of the system being in the clean limit, a property further 
confirmed by our detailed analysis of the frequency dependent penetration depth. The
normal-state optical conductivity can be parameterized by a Drude-Lorentz
dielectric function. We find that all parameters in the
Drude-Lorentz model are temperature independent, except for the
scattering rate of a coherent Drude peak, representative of
quasiparticles on the electron Fermi sheets. A multiband analysis of the
scattering rate indicates a non-Fermi-liquid
behavior. The linear behavior observed for the scattering rate
is compatible with spin fluctuations
and supports the presence of a quantum critical point.

%%%%%%%%%%%%%%%%%%%%%%%%%%%%%%%%%%%%%%%%%%%%%%%%%%%%%%%%%%%%%%%%%%%%%%%%%%%%%%%
%
% Acknowlegment
%
\begin{acknowledgments}
Part of this work was supported by HLD at HZDR, member of
the European Magnetic Field Laboratory (EMFL).
\end{acknowledgments}
%
% The bibliography (BibTeX)
%
\bibliography{biblio}

\end{document}